\documentclass[prb,twocolumn,superscriptaddress,showpacs,amsmath,amssymb]{revtex4}

\usepackage{bm}

\usepackage{graphicx}
\usepackage{latexsym}
\usepackage{amsmath}
\usepackage{amssymb}
\usepackage{amsfonts}
\usepackage{color}
\usepackage{bm}
\usepackage{verbatim}

\begin{document}

\title{Magnetic resonance within vortex cores in the B phase of superfluid $^3$He.}

\author{I.M. Khaymovich}
 \affiliation{Institute for Physics of Microstructures RAS, 603950
Nizhny Novgorod, Russia.}

\author{M.\,A.\,Silaev}%
 \affiliation{Institute for Physics of Microstructures RAS, 603950
Nizhny Novgorod, Russia.}
 \affiliation{Low Temperature Laboratory, Aalto University, P.O. Box 15100, FI-00076 AALTO, Finland}


\date{\today}

\begin{abstract}
We investigate a magnetic susceptibility of vortices in the B
phase of multicomponent triplet superfluid $^3$He focusing on a
contribution of bound fermionic states localized within vortex
cores.
 Several order parameter configurations relevant to different types of quantized
vortices in $^3$He B are considered. It is shown quite generally
that an ac magnetic susceptibility has a sharp peak at the
frequency corresponding to the energy of interlevel spacing in the
spectrum of bound fermions. We suggest that measuring of a
magnetic resonance within vortex cores can provide a direct probe
of a discrete spectrum of bound vortex core excitations.
\end{abstract}


 \maketitle



\section{Introduction.}

\begin{figure}[h!]
 \centerline{\includegraphics[width=1.0\linewidth]{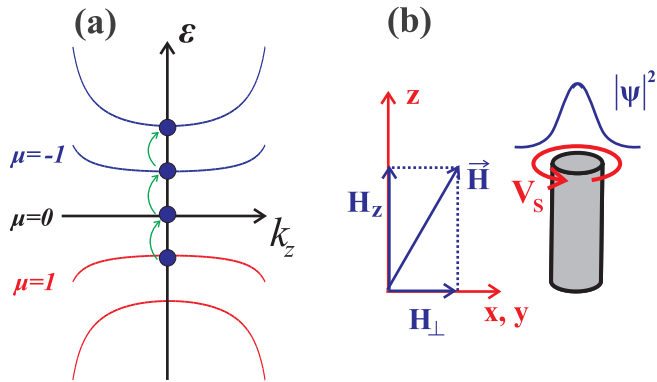}}
 \caption{\label{Scetch}
 (a) Sample
 spectrum of Caroli-de Gennes-Matricon states for vortex in $p$-wave
 superfluid.
 (b) Sketch of the system
 consisting of vortex line directed along the z axis under the action of magnetic field.
  The time dependent magnetic field ${\bf H}={\bf H} (t)$ leads to the
 transitions in the ladder of Caroli-de Gennes-Matricon states.
 The interlevel transitions caused by the time-dependent magnetic field
 are shown schematically in the panel (a) by arrows.}
   \end{figure}

Probing of a quasiparticle spectrum in Fermi superfluids
containing quantized vortices has been a challenging problem since
the pioneering work of Caroli, de Gennes and Matricon \cite{CdGM}.
They predicted theoretically that the internal electronic
structure of
 quantized vortices in superconductors should
consist of low energy fermionic excitations localized within the
vortex cores with characteristic interlevel
 spacing defined as $\Delta_0^2/E_F\ll\Delta_0$, where
 $\Delta_0$ is the energy gap far from the vortex line and $E_F$ is the Fermi energy.
 For conventional s-wave superconductors with axially symmetric vortex lines
 the low-energy ($|\varepsilon|\ll\Delta_0$) part of the quasiparticle spectrum has the following form
 \begin{equation}\label{CdGMspectrum}
  \varepsilon(\mu,k_z)\simeq-\hbar\omega_v\mu.
 \end{equation}
 Here $\mu$ and $k_z$ are the projections of quasiparticle angular and linear
 momenta onto the vortex axis and
 \begin{equation}\label{omega(kz)}
  \hbar\omega_v\approx\frac{\Delta_0}{\sqrt{k_F^2-k_z^2}\xi},
 \end{equation}
 where $\xi$ is a coherence length in superconductor or
 superfluid.
 Due to the quantization of angular motion the projection of
 angular momentum $\mu$ takes discrete values which are half
 integer for the $s$ wave superconductor and integer for the $p$
 wave superconductors and superfluids like $^3$ He (see
 Ref.\onlinecite{VolovikPwave}). Thus the spectrum of bound vortex
 core fermions (\ref{CdGMspectrum}) consists of a ladder of states which varies from
 $\Delta_0$ at $\mu=-\infty$ to $\Delta_0$ as $\mu=\infty$. An example of the
  Caroli - de Gennes - Matricon spectrum for the vortex in $p$-wave superfluid
   is shown in Fig.\ref{Scetch}a.

Bound fermions determine many of the physical properties of
vortices in superconductors and Fermi superfluids like $^3$He. In
particular, vortex dynamics is determined by the kinetics of
vortex core quasiparticles (see for example review
Ref.\onlinecite{KopninReview} and also
Refs.\onlinecite{KopninSalomaa, VolovikSpFlow,Stone}) and low
temperature thermodynamic properties are determined by the
peculiarities of spectrum of bound fermions. Recently there has
been much attention focused on the vortex core states in chiral
triplet superconductors in connection with the topologically
protected zero energy states being a realization of
self-conjugated Majorana fermions (see for example
Ref.\onlinecite{Majorana}). Such objects have topological nature
\cite{VolovikPwave} and are the key ones for the realization of
topological quantum computation which is presently generating much
interest\cite{TopologicalComputing}.

 One of the most striking
demonstrations of the existence of such bound states have been
tunneling spectroscopy (STM) experiments showing a peak of the
local density of quasiparticle states for the scanning tip
position at the vortex center \cite{STM}. However the spatial and
energy resolution of STM techniques has been insufficient to
investigate a discrete energy levels of vortex core quasiparticles
which were predicted by Caroli, de Gennes and Matricon.

Another experimental tool to study the discreet nature of the
vortex core quasiparticle spectrum would be measuring of the
dynamical responses of Fermi superfluids determined by the motion
of vortices. In particular the ac conductivity of superconductors
in superclean regime should demonstrate a peak at the frequency
determined by the interlevel transitions of quasiparticles
localized within the vortex cores \cite{KopninKravtsov}. The same
is also true for the drag viscosity coefficient in the B phase of
$^3$He (see Refs.\onlinecite{KopninReview,KopninSalomaa}). Of
particular interest are the effects of fermion bound states in a
vortex core on the rotational dynamics of vortices with
spontaneously broken axisymmetry\cite{RotatingVortex}. It has been
predicted that a resonance
 absorption of an external rf field can occur
  at the external frequency comparable with the interlevel distance of localized states
which is similar to the cyclotron Landau damping.

 However  experimental
realization of the mechanisms described above is still lacking. In
superconductors the main obstacle is a rather restricting
condition for the material purity, which is hardly obtainable in
conventional superconductors. In superfluid $^3$He the reason is
an extremely high frequency of vortex motion which is required to
excite the transitions between quantized energy levels inside
vortex cores. Thus it seems that probing of the Caroli- de Gennes
- Matricon theory with the help of existing to date experimental
approaches is unattainable.



Our basic idea is to employ for this purpose the Zeeman
interaction of quasiparticles in Fermi systems with external
magnetic field. Quasiparticle spin degree of freedom has been paid
little attention up to date in connection with the properties of
the quasiparticle spectra of vortex cores. The reason for it is
that in conventional superconductors with singlet pairing the spin
is a completely independent variable and it does not affect the
orbital motion of quasiparticles. On the other hand the situation
is completely different in multicomponent triplet Fermi
superfluids. Indeed, in this case the Cooper pair wave function is
a superposition of different spin states:
\begin{equation}\label{Dvector}
\hat\Delta=-i ({\bf \hat \sigma}\cdot{\bf d})\hat \sigma_y,
\end{equation}
 where ${\bf d}$ is a vector in 3D space
and $\sigma_{x,y,z}$ are Pauli matrices in conventional spin
space. In case if the direction of spin vector ${\bf d}$ is
different at the different points of Fermi sphere, quasiparticles
"see" the direction of Cooper pair spin which essentially depends
on the direction of quasiparticle wave vector ${\bf k}$. If there
is no other spin dependent potentials then quasiparticle spin is
determined only by the order parameter and it occurs to be
dependent on the direction of quasiparticle propagation. Thus we
may conclude that in multicomponent triplet superfluids
quasiparticle spin and orbital degrees of freedom are effectively
coupled.

 The schematic configuration of the system that we propose to study is
shown in Fig.\ref{Scetch}b. In general we consider vortices in the
B phase of superfluid $^3$He under the action of external magnetic
field. We assume that the magnetic field applied to the vortex
consists of a large constant component along the vortex axis $H_z$
and a small time-varying component directed perpendicular to the
vortex axis $H_\perp=H_\perp (t)$. Is we will show below due to
the effective spin - orbital coupling for quasiparticles in $^3$He
B the time dependent magnetic field can excite the transitions in
a ladder of Caroli-de Gennes- Matricon states through the Zeeman
interaction with quasiparticle spins. These transitions are shown
schematically by arrows in Fig.\ref{Scetch}a. In turn, such
transitions lead to the resonant energy absorption at the
frequency determined by the interlevel spacing which observation
would be the demonstration of a discrete structure of spectrum of
bound vortex core states (\ref{CdGMspectrum}).

 It is natural to expect
 that the interlevel energy spacing which in $^3$He is of the order
 $\omega_v \sim 0.1\; {\rm MHz}$ (see Ref. \onlinecite{KopninSalomaa}) would determine the resonant frequency of
  magnetic response of vortex cores. Note that such resonance
 is qualitatively different from the magnetic resonance of
 free nuclear spins (NMR) when the resonant frequency is determined by the
 constant component of applied magnetic field. Within the frequency domain and range of magnetic fields
 we are interested in such type of "usual" NMR can be
 neglected. Indeed, the gyromagnetic ratio of $^3$He nucleus is of the order
 $\gamma_{g}\sim  10^4\; {\rm Hz/G}$
 therefore the Larmour frequency of $\omega_L=\gamma_{g} H \sim 0.1\; {\rm MHz}$
  corresponds to magnetic field of about $H=\omega_L/\gamma_g \sim 10\; {\rm G}$
  which is much less than typical magnetic
 fields of the order $H\sim 100\; {\rm G}$ used in NMR experiments in superfluid
 $^3$He. Under such conditions the frequencies that we will be interested in lie outside the region
 of usual NMR peak $\omega_v\ll \omega_L$ which means that we can
 completely neglect the Larmour spin precession.

This paper is organized as follows. In Sec.~\ref{BasicEq} we give
an overview of the theoretical framework which is employed in this
work, namely the Bogoliubov--de Gennes theory which we use to
analyze the spectra of bound fermions and kinetic theory to
calculate a nonequilibrium magnetization of vortex cores in time
dependent magnetic field. The main results are presented in
Sec.~\ref{Results}, in particular the transformation of the
quasiparticle spectra is discussed in Secs.~\ref{Results:spectra1}
and ~\ref{Results:spectra2}. The nonequlibrium magnetization and
paramagnetic susceptibility of vortex cores and dissipation losses
are addressed in Sec.~\ref{Results:magn}. We summarize our results
in Section~\ref{sec:summary}. Some of the details of our
calculations are given in appendices.


\section{Basic equations.} \label{BasicEq}


\subsection{Order parameter.}

In general order parameter of triplet superfluid has the form
given by Eq.(\ref{Dvector}). Further assuming a $p$ - wave pairing
which is the most relevant case for the superfluid $^3$He we write
the order parameter in the form
\begin{equation}\label{OPgen}
 \hat\Delta_{\bf k}=A_{\alpha i} (i{\bf \hat
 \sigma}_\alpha\hat\sigma_y)(k_i/k_F).
 \end{equation}
  The $3\times 3$  matrix with complex
 coefficients $A_{\alpha i}$ in the expression above can be
 represented as an expansion
\begin{equation}\label{ExpansionA}
 A_{\alpha i}=\Delta \sum_{\mu\nu} a_{\mu\nu}\lambda^\mu_\alpha
 \lambda^\nu_i,
 \end{equation}
where $\lambda^{\pm}_{i,\alpha}=({\bf x}_{i,\alpha}\pm i {\bf
y}_{i,\alpha}
 )$,
 $\lambda^0_{i,\alpha}={\bf z}_{i,\alpha}$
 are the eigenfunctions of orbital momentum $\hat L_z$ and spin $\hat S_z$ with
 eigenvalues $\mu$ and $\nu$.

 Further in this paper we will deal with axially symmetric quantized vortices
 in superfluid $^3$He.
The axial symmetry of such objects is described by a generator of
rotations around the vortex axis $\hat Q=\hat J_z-M\hat I$, where
$\hat J_z=\hat L_z+\hat R\hat S_z$ is the projection of the
internal angular momentum of Cooper pairs onto the vortex. We
introduce here the operator of 3D rotation $\hat R$ which
transforms the coordinate axes in spin space into the ones in
orbital space. Such rotation of coordinate axes leads to the
transformation of the order parameter according to the following
rule:
\begin{equation}\label{RotateOP}
 \tilde{A}_{\alpha i}=R_{\alpha\beta}A_{\beta i}.
\end{equation}
As we will see further the relative rotation of spin and orbital
coordinate axes does not lead to qualitatively new results.
Therefore basically we will assume that the spin and orbital
quantization axes coincide with the vortex axis $z$ and will just
briefly discuss the appropriate changes in the resulting formula
which take into account the rotation (\ref{RotateOP}).

The axial symmetry of the order parameter components satisfying
the equation $\hat Q A_{\alpha i}=0$ is described by the following
choice of the coefficients in the expansion (\ref{ExpansionA})
\begin{equation}\label{ExpansionAcoeff}
 a_{\mu\nu}=C_{\mu\nu}(r) e^{i(M-\mu-\nu)\varphi},
 \end{equation}
  where $\varphi,r$ are polar coordinates with the origin at the
  vortex center and the dependencies of the order parameter components on the radial
 distance $r$ are given by the functions $C_{\mu\nu}(r)$.

 The coefficient $M$ in Eq.(\ref{ExpansionAcoeff}) is determined by the vorticity $M_v$ and the
internal angular momentum of Cooper pairs far from the vortex
axis.  For example the B phase of $^3$He is characterized by the
zero internal angular momentum and therefore $M=M_v$. However in
the A phase of $^3$He as well as in the chiral state of triplet
superconductor $Sr_2RuO_4$ in the homogeneous state, we have
$J_z=\pm 1$ and therefore $M=M_v\pm 1$.

Let us now consider the transformation of the vortex order
parameter under the action of several discrete symmetries: time
inversion $T$, space inversion $P$ and rotation by the angle $\pi$
over the $x$ axis $U_2$. Under the time inversion we get $T
A_{\alpha i}=A^*_{\alpha i}$ therefore
 \begin{equation}\label{T-transform}
 T(C_{\mu\nu})=C^*_{-\mu,-\nu}.
 \end{equation}
Under the spatial inversion we get
 \begin{equation}\label{P-transform}
  P(C_{\mu\nu})=(-1)^{M-\mu-\nu}C_{\mu\nu}
 \end{equation}
 and finally
 \begin{equation}\label{U-transform}
 U_2(C_{\mu\nu})=(-1)^{\mu+\nu}C_{-\mu,-\nu}.
 \end{equation}

As we will see later it is very important that we can construct
 a general expression for the pseudoscalar from the
amplitudes of the order parameter components $C_{\mu\nu}$. Indeed,
let us denote
 $$
 \tilde{C}_{n}=\sum_{|\mu+\nu|=n}C_{\mu\nu}.
 $$
 Note that $n=0,1,2$ since $|\mu|=|\nu|=1$ in $^3$He.
   Then a general expression of the pseudoscalar which can be composed
   from the order parameter components has the following gauge invariant form:
 \begin{equation}\label{pseudoscalar}
  \alpha_{p}=Im(\tilde{C}^*_{0}\tilde{C}_{1})+Im(\tilde{C}^*_{2}\tilde{C}_{1}).
 \end{equation}
 From Eqs.(\ref{T-transform},\ref{P-transform},\ref{U-transform}) it is obvious that
 $$
 T(\tilde{C}_{n})=\tilde{C}^*_{n},
 $$
 $$
 P(\tilde{C}_{n})=(-1)^{n+M}\tilde{C}_{n},
 $$
 $$
 U_2(\tilde{C}_{n})=(-1)^n\tilde{C}_{n}.
 $$
 Therefore assuming that the vortices are singly quantized $M=\pm 1$
 from the above expressions we get that
 $$
 P,T,U_2(\alpha_{p})=-\alpha_{p}.
 $$

In this paper we will consider only the vortices in the B phase of
superfluid $^3$He. For singly quantized vortices $M=\pm 11$ there
can exist five basic components of the order
 parameter. Among them are $C_{1,-1}$, $C_{-1,1}$ and $C_{00}$
  which correspond to the main B phase, $C_{0,1}=C_{A}$ and
$C_{1,0}=C_{\beta}$ which correspond to the
 additional A and $\beta$ phases localized inside vortex core. The additional A phase
 has a zero spin projection ($\mu=0$) and unit projection of
 orbital momentum ($\nu=1$) on the $z$ axis while $\beta$ phase has $\mu=1$ and $\nu=0$
 \cite{VolovikRMP}.
  Far from the vortex core at $r\gg\xi_v$ only B superfluid phase
 exists so that $C_{1,-1}=C_{-1,1}=C_{00}=1$
  and $C_{A,\beta}=0$. The vortex type is determined by the behaviour of amplitudes $C_{\mu\nu}$
  at smaller distances $r\sim\xi_v$ and there exist five types of vortices \cite{VolovikRMP}.

Vortices of $o$ and $u$ types are singular so
  that only the superfluid components of B phase $C_{1,-1}$, $C_{-1,1}$ and $C_{00}$
  are nonzero.
   These amplitudes are real for the most symmetric $o$ vortex which is invariant under the
   action of three basic discrete symmetry transformations: $P_1=P$, $P_2=PTU_2$ and
   $P_3=TU_2$. The less symmetric $u$ vortex with conserved parity $P_1=P$ but
 broken $P_3=TU_2$ discrete symmetry is characterized by the
 complex amplitudes of the order parameter components.

Nonsingular $v$, $w$ and $uvw$ vortices have superfluid cores with
the inclusion of A and $\beta$ phases.
 The functions $C_{A,\beta}=C_{A,\beta}(r)$ describing the spatial distributions of additional A and $\beta$
  phases inside vortex core are finite at $r=0$ and vanish outside the core at $r\gg\xi_v$.
 The $v$ and $w$ vortices are characterized by real B phase
 amplitudes. If $C_{A,\beta}$ are also real
  then we have a $v$ vortex with conserved $P_2=PTU_2$ symmetry. The case when $Re(C_{A,\beta})=0$,
  $Im(C_{A,\beta})\neq 0$ corresponds
 to $w$ vortex with conserved $P_3=TU_2$ symmetry. The less
 symmetric $uvw$ vortex with all discrete symmetries $P_1,P_2,P_3$ broken
 has complex amplitudes of B, A and $\beta$ phases.

Returning to the definition of the pseudoscalar
(\ref{pseudoscalar}) and applying it to the particular case of
vortices in $^3$He B it is easy to see that $\alpha_p=0$ for the
singular $o$ and $u$ vortices as well as for the nonsingular $v$
vortex. At the same time for $w$ and $uvw$ vortices we get a
non-zero pseudoscalar $\alpha_p\neq 0$. Assuming for example the
model situation when $C_{1,-1}=C_{-1,1}=C_{00}=C_B (r)$ and
$C_A\neq 0$ we obtain from Eq.(\ref{pseudoscalar}) that for $w$
and $uvw$ vortices
\begin{equation}\label{pseudoscalarUVW}
\alpha_p=Im (C^*_B C_A).
\end{equation}
 As we will see below such classification of vortices in terms of the
pseudoscalar has a direct connection with qualitatively different
modification of quasiparticle spectrum by an external magnetic
field.


\subsection{Quasiclassical Bogoulubov-de Gennes equations.}

 Let us now turn to the spectrum of
quasiparticles, which is described by the Bogoulubov- de Gennes
(BdG) equations. The quasiclassical form of BdG equations then
reads as follows:
 \begin{equation}\label{BdG-U}
 -i\frac{\hbar k_\perp}{m} \frac{\partial}{\partial s}
 U+\hat\Delta_{\bf k} V=\left(\varepsilon-\hat P\right)U,
 \end{equation}
 \begin{equation}\label{BdG-V}
 i\frac{\hbar k_\perp}{m} \frac{\partial}{\partial s}
 V+\hat{\Delta}^+_{\bf k}U=\left(\varepsilon+\hat P^*\right)V.
 \end{equation}
  Here $k_\perp=\sqrt{k_F^2-k_z^2}$ is the component of
  Fermi momentum perpendicular to vortex axis ${\bf z}$,
   $\hat P= \mu_B ({\bf H}\cdot{\bf \hat \sigma})$ is the Zeeman
   term and $\hat \Delta$ is a gap operator (\ref{OPgen}).
    We assume that the magnetic field ${\bf H}$ applied to
the system enters the equations only through the Zeeman terms and
in general the magnetic field dependence of gap function $\hat
\Delta_{\bf k}=\hat \Delta_{\bf k} ({\bf H})$.

   Within quasiclassical formalism we should express the
   real space coordinates through the coordinate
\begin{equation}\label{Scoordinate}
s=({\bf n_k}\cdot {\bf r})
\end{equation}
 along the trajectory characterized by the direction of quasiparticle
 momentum ${\bf n_k}= {\bf k_\perp}/k_\perp= (\cos\theta_p, \sin\theta_p)$ and the
   impact parameter
\begin{equation}\label{Imparameter}
b=({\bf z}\cdot [{\bf n_k}\times {\bf r}]).
\end{equation}
The impact parameter is related to the angular momentum
projections along the vortex axis $\mu$ through the usual
classical mechanics formula $\mu=-k_\perp b$.

The energy spectrum of the quasiclassical BdG equations
$\varepsilon=\varepsilon (\mu,\theta_p,k_z)$ depends on trajectory
angle $\theta_p$, angular momentum $\mu$ and linear momentum $k_z$
projections. Later we will use the following general symmetry of
quasiclassical spectrum (see Appendix A)
 \begin{equation}\label{QuasiSymmetry}
  \varepsilon(\mu,\theta_p,k_z)=-\varepsilon(-\mu,\theta_p+\pi,-k_z),
 \end{equation}
 which holds for an arbitrary triplet order parameter.
 Also we will assume for simplicity that the order parameter is
 symmetric with respect to the sign change of magnetic field
 $$
 \hat \Delta_k({\bf H}) =\hat \Delta_k (-{\bf H})
 $$
 which is justified as long as we neglect the spontaneous
 magnetic moment of vortex cores \cite{FerromagneticCore}.
 Under such assumption the following symmetry of the spectrum is
 valid (see Appendix A):
 \begin{equation}\label{QuasiSymmetryH}
  \varepsilon({\bf H})=\varepsilon(-{\bf H}).
 \end{equation}

 Let us now consider
 the Green's functions of BdG system of equations
 (\ref{BdG-U},\ref{BdG-V})
\begin{equation}\label{Green}
 \check{G}^{R(A)}(s_1,s_2)=\sum_n\frac{\Psi_n^+(s_1) \Psi_n (s_2)}{\epsilon-\varepsilon_n\pm
 i\delta},
 \end{equation}
  where we introduce the two component eigenfunctions $\Psi_n=(U_n,V_n)$ of BdG
 equations corresponding to the energy level $\varepsilon_n$.
 Further we will use a relation between the functions (\ref{Green}) and
 the quasiclassical Green's functions $\check{g}^{R(A)}=\check{g}^{R(A)} ({\bf
 k},{\bf r},\epsilon)$ having the following form:
 $$
 \check{g}^{R(A)}(s)=-i\frac{\hbar^2
 k_\perp}{m}\left[\check{G}^{R(A)}(s,s+0)+\check{G}^{R(A)}(s,s-0)\right],
 $$
 where the coordinate $s$ along trajectory is related to the vectors
 ${\bf r}$ and ${\bf k}$ according to Eq.(\ref{Scoordinate}).

 Then taking the derivatives of the both sides of BdG system
 and using the normalization condition $\int \Psi_n \Psi^+_n ds
 =1$ it is easy to obtain the expression which we will use later:
 \begin{equation}\label{usefulEq}
  \int Tr\frac{\partial\check{H}}{\partial {\bf H}}
 (\check{g}^{R}-\check{g}^{A}) ds= 4\pi\frac{\hbar^2 k_\perp}{m}\sum_n
 \delta(\epsilon-\varepsilon_n) \frac{\partial \varepsilon_n}{\partial{\bf
 H}}.
 \end{equation}
 In the equation above we use the following magnetic field dependent part of the
 hamiltonian
\begin{equation}\label{Hamiltonian}
 \check {H}=\begin{pmatrix}
   \mu_B ({\bf \sigma}\cdot {\bf H}) & \Delta_{\bf k} \\
   \Delta^+_{\bf k} & -\mu_B ({\bf \sigma}^*\cdot {\bf H}) \
 \end{pmatrix},
 \end{equation}
 where the diagonal terms define the interaction of nuclear spins
 with the external magnetic field and $\mu_B$ is a nuclear
 magneton. This hamiltonian can be rewritten with the help of spin
 operator introduced above $
 \check {H}=\mu_B ({\bf \check{S}}\cdot{\bf H})+\check{\Delta}_{\bf k}$,
 where we have introduced a matrix gap function
 $$
 \check{\Delta}_{\bf k}=\begin{pmatrix}
   0 & \hat\Delta_{\bf k} \\
   \hat\Delta^+_{\bf k} & 0 \
 \end{pmatrix}
 $$
 and the operator of quasiparticle spin
\begin{equation}\label{SpinNambu}
{\bf \check{S}}=(\hat\tau_3 \hat\sigma_x, \hat\sigma_y,\hat\tau_3
\hat\sigma_z).
\end{equation}
Here the Pauli matrices $\hat\sigma_{x,y,z}$ and
$\hat\tau_{1,2,3}$ act in spin and Nambu spaces correspondingly.
Note that the Eq.(\ref{usefulEq}) takes into account not only the
energy shift due to the interaction of nuclear spin with magnetic
field but also the field dependence of gap function $\Delta_{\bf
k}=\Delta_{\bf k} ({\bf H})$. In the next subsection we will use
the relation (\ref{usefulEq}) to derive the expression for
magnetization of vortex cores.


\subsection{Magnetization of vortex cores and energy dissipation.}

 At first we need
to derive an expression for the paramagnetic response of vortex
cores in superfluid $^3$He B. Our essential interest is in the
dynamics of vortex core magnetization driven by the time-dependent
external magnetic field and the corresponding energy dissipation
due to the interaction of quasiparticles with the heat bath.

We start with the exact expression of spin magnetic moment valid
for the general non-equilibrium system:
\begin{equation}\label{MagnExact}
{\bf M}= -\mu_B \nu_0 \int \frac{d\epsilon}{4} \frac{d\Omega_{\bf
k}}{4\pi} d {\bf r}\; Tr\; {\bf \check{S}}\check{g}^K,
\end{equation}
where we introduce the quasiclassical Keldysh function
$\check{g}^K=\check{g}^K ({\bf k},{\bf r}, t, \epsilon)$ and the
operator of quasiparticle spin is given by (\ref{SpinNambu}). The
integration in momentum space is done over the Fermi sphere so
that $(d\Omega_{\bf k})/(4\pi)=(dk_zd\theta_p)/(4\pi k_F)$ and
$\nu_0$ is a density of states at the Fermi level.

 In general the average rate of dissipation losses in the system can be calculated as the work
 of the source of external magnetic field as follows
 \begin{equation}\label{DissipationGeneral}
 Q_\varepsilon=-\langle {\bf M}\cdot {\bf \dot{H}}\rangle_t,
 \end{equation}
where the brackets denote the time average. However some parts of
the total magnetization (\ref{MagnExact}) of the system correspond
to a reversible exchange of the energy between the source and the
system and for the time periodical processes drop out from the
expression for the work of external source.

To take into account such reversible energy exchange let us note
at first that the average rate of energy loss does not change if
we add to Eq.(\ref{DissipationGeneral}) the time averaged change
of the system energy, or to be more precise the thermodynamic
potential $\Omega$. Indeed, for the time periodical processes the
average change of system thermodynamic potential is zero. Thus we
can substitute the expression for the dissipation losses
(\ref{DissipationGeneral}) by the following
\begin{equation}\label{DissipationGen}
Q_\varepsilon=-\langle \frac{\delta E_{tot}}{\delta {\bf H}}\cdot
 {\bf\dot{H}}\rangle_t.
\end{equation}
In the expression above we introduce the total energy which
describes the interaction between the vortex and the source of
external magnetic field $E_{tot}=\Omega+E_m$, where $\Omega$ is
the thermodynamic potential of the system in the magnetic field
and $E_m$ is the part of magnetic energy which describes the work
done by the external source $\delta E_m=-\delta R= {\bf
M}\cdot\delta{\bf {H}}$. The variation of $E_{tot}$ with respect
to the magnetic field
\begin{equation}\label{Eint-Var}
  \frac{\delta E_{tot}}{\delta {\bf H}}= \frac{\delta \Omega}{\delta {\bf
  H}}+{\bf M}
\end{equation}
   gives exactly the flow of the energy to
the heat bath, i.e. the dissipation losses.

To evaluate the expression for the energy dissipation
(\ref{DissipationGen}) we should calculate the variation of
thermodynamic potential with the magnetic field
$\delta\Omega/\delta{\bf H}$. For this purpose we use the standard
expression for the variation of the thermodynamic potential [see
for example Ref.\onlinecite{Kopnin}]:
 $$
 \delta\Omega=\nu_0\int\; Tr\; [\delta\check{H} \check{g}^{(st)} ({\bf k},{\bf
 r})]\frac{d\epsilon}{4} \frac{d\Omega_{\bf k}}{4\pi} d {\bf r}
 $$
\begin{equation}\label{VarThPot}
- \int \; Tr\; \frac{\delta\left(\Delta_{\bf k}({\bf
r})\Delta^*_{\bf k}({\bf r})\right)}
 {|\lambda|}\frac{d\Omega_{\bf k}}{4\pi} d {\bf
 r},
 \end{equation}
  where $\lambda$ is the weak coupling constant and
   the variation of hamiltonian (\ref{Hamiltonian}) is given by
 \begin{equation}\label{Hamiltonian-Var}
 \delta \check {H}=\left[\mu_B {\bf \check{S}}+\frac{\partial\check{\Delta}_{\bf
 k}}{\partial{\bf H}}\right] \cdot \delta{\bf H}.
 \end{equation}
 In Eq.(\ref{VarThPot}) we have
 introduced the stationary
 part $\check{g}^{(st)}$ of the total Green's function $\check{g}^{K}$
 which is time independent and can be expressed
 through the stationary retarded and advanced Green functions
 $\check{g}^{R(A)}_\epsilon ({\bf k},{\bf r})$ and the equilibrium
 distribution function $f^{(0)}(\epsilon)=\tanh(\epsilon/2T)$ as
 follows:
 $$
 \check{g}^{(st)}=(\check{g}^{R}_\epsilon-\check{g}^{A}_\epsilon)f^{(0)}(\epsilon).
 $$

 When calculating the variation of thermodynamic potential
 we should take into account the self-consistent change of the
 gap function determined by the following equation:
 \begin{equation}\label{SelfConsistency}
 \Delta_{\bf k} ({\bf r},t)=\lambda \nu_0\int \frac{d\epsilon}{4} \frac{d\Omega_{\bf k_1}}{4\pi} V({\bf k},{\bf
 k_1}) f^K ({\bf k_1},{\bf r},t),
 \end{equation}
 where $V({\bf k},{\bf k_1})$ is a pairing interaction which for the p-wave superfluid
 has the form $V({\bf k},{\bf k_1})=({\bf k}\cdot{\bf k_1})$. Note
 that it is a total non-equilibrium Green's function $f^K ({\bf k},{\bf r},t)$ which enters the
 self consistency equation (\ref{SelfConsistency}).

 Substituting the Eq.(\ref{SelfConsistency}) into the expression
 for the variation of thermodynamic potential (\ref{VarThPot}) we
 immediately obtain that
  $$
 \frac{\delta\Omega}{\delta {\bf H}}=\mu_B\nu_0\int\; Tr\; [{\bf\check{S}} \check{g}^{(st)} ({\bf k},{\bf
 r})]\frac{d\epsilon}{4} \frac{d\Omega_{\bf k}}{4\pi} d {\bf r}
 $$
\begin{equation}\label{VarThPot-1}
-\nu_0\int\; Tr\; \left[\frac{\check{\partial \Delta}_{\bf
k}}{\delta {\bf H }} \check{g}^{(nst)} ({\bf k},{\bf
 r})\right]\frac{d\epsilon}{4} \frac{d\Omega_{\bf k}}{4\pi} d {\bf r},
 \end{equation}
 where we have introduced a non-stationary part of Green's
 function as
\begin{equation}\label{GreenNst}
 \check{g}^{(nst)}=\check{g}^{K}-\check{g}^{(st)}.
\end{equation}
 The next step is to get use of the
 Eqs.(\ref{MagnExact},\ref{VarThPot-1}) substituting them to the Eq.(\ref{Eint-Var}).
Then we obtain the expression for magnetization (\ref{MagnExact})
  to calculate the variation of interaction energy and obtain that
  $(\delta E_{tot})/(\delta {\bf H})={\bf M}_{qp}$ where
 \begin{equation}\label{MagnQP}
   {\bf M}_{qp}=
   -\nu_0 \int \frac{d\epsilon}{4} \frac{d\Omega_{\bf k}}{4\pi} d {\bf r}\; Tr\;
  \frac{\partial\check{H}}{\partial {\bf H}}\check{g}^{(nst)}.
  \end{equation}

 The introduced quantity ${\bf M}_{qp}$ has the physical meaning
 of magnetization of the ensemble of quasiparticles which reside
 within vortex core. Indeed for the magnetization of an arbitrary
 system of non-interacting particles described by a single -
 particle hamiltonian $\check{H}$ the operator of magnetic moment
 has the form\cite{White} $\check {\bf M}_{qp}= -(\partial \check{H})/(\partial {\bf H}) $
 and the expression (\ref{MagnQP}) yields the non-stationary part
 of the thermodynamic average of $\check {\bf M}_{qp}$.

 Further we will deal with the linear magnetic response and monochromatic
 processes so that it is convenient to introduce a paramagnetic susceptibility
 of quasiparticles in the frequency domain $\hat\chi_{qp}=\hat\chi_{qp} (\omega_{rf})$ as follows
 $$
 {\bf M}_{qp} (\omega_{rf})=\hat\chi_{qp} {\bf H} (\omega_{rf}),
 $$
 where
 $$
 {\bf H}(\omega_{rf})=\int_{-\infty}^{\infty} {\bf H} (t) e^{i\omega_{rf} t} dt,
 $$
 $$
 {\bf M}_{qp} (\omega_{rf})=\int_{-\infty}^{\infty} {\bf M}_{qp} (t) e^{i\omega_{rf} t}
 dt.
 $$
 Then for the monochromatic magnetic field
 the dissipation losses are determined by the standard
 expression
\begin{equation}\label{DissLossSuscept}
 Q_\epsilon= \frac{\omega_{rf}}{2} \langle {\bf H}^* \hat
\chi^{\prime\prime}_{qp}  {\bf H}
 \rangle_t,
 \end{equation}
 where $ \hat\chi^{\prime\prime}_{qp}= Im \hat\chi_{qp}$.

The expression (\ref{MagnQP}) has an essential advantage since it
allows to take into account the dependence of the energy gap on
the magnetic field $\Delta_{\bf k}=\Delta_{\bf k}({\bf H})$, which
in general can not be neglected.
 As we will see below the only thing we should know to calculate ${\bf
M}_{qp}$ according to Eq.(\ref{MagnQP}) is a dependence of
quasiparticle energy on the magnetic field. The quasiparticle
spectrum can be either calculated exactly for model situations
with non-self consistent gap $\Delta_{\bf k}$ or it can be taken
in some general form determined by the symmetry of the system. We
will follow the latter way since it allows to take into account
the magnetic field dependence of the gap function without
extensive self-consistent calculations.


\subsection{Distribution function and kinetic equation.}

To proceed further we assume that the deviations from equilibrium
are small and the rate of magnetic field variation is much slower
than the relaxation time of the gap function. Then we use the
approximate expression for the non-stationary part of the Keldysh
function valid for the slow variation of the system parameters
\cite{Kopnin}:
\begin{equation}\label{gNst}
 \check{g}^{(nst)}=(\check{g}^{R}-\check{g}^{A}) f_1,
 \end{equation}
 where the spectral Green functions $\check{g}^{R(A)}$ are taken for the stationary
 system and the function $f_1=f_1({\bf k},{\bf r},t)$ determines
 the non-equilibrium deviation of the symmetric part of generalized distribution
 function.

Our goal now is to rewrite the Eq.({\ref{MagnQP}}) in terms of the
 spectrum of quasiclassical BdG equation. At first let us use the new coordinate system
defined by the relations (\ref{Scoordinate},\ref{Imparameter}).
Then the integration over $d {\bf r}d\Omega_{\bf k}/(4\pi)$ in
Eq.({\ref{MagnQP}}) transforms to the integration over
$d\theta_p\; d\mu\; ds\; dk_z/(4\pi k_Fk_\perp)$.
  Substituting the expression (\ref{gNst}) into Eq.(\ref{MagnQP})
  and using the relation (\ref{usefulEq}) we finally obtain the following
equation for the non-stationary magnetization
\begin{equation}\label{MagnQPst}
{\bf M}_{qp}=-\frac{1}{8\pi^2} \sum_n \int d\mu d\theta_p d k_z
\frac{\partial\varepsilon_n}{\partial {\bf H}}f_1.
\end{equation}

Finally to calculate the magnetization with the help of
Eq.(\ref{MagnQPst}) we only need to know the quasiparticle
spectrum $\varepsilon_n=\varepsilon_n(\mu,\theta_p)$ which can be
found solving the BdG equations (\ref{BdG-U},\ref{BdG-V}) and the
distribution function
$f(\mu,\theta_p,t)=f_0(\varepsilon_n(\mu,\theta_p))+f_1$, which
obeys the kinetic equation\cite{Kopnin}:
\begin{equation}\label{kinetic}
 \frac{\partial f}{\partial t}+ \frac{\partial f}{\partial
 \theta_p}\dot{\theta_p}+
 \frac{\partial f}{\partial \mu}\dot{\mu}=St(f).
\end{equation}
  The collision integral in the right hand size of Eq.(\ref{kinetic})
 can be taken in the model relaxation time approximation:
 $St(f)=(f-f_0)/\tau$, where $f_0$ is a equilibrium function which
 we take in the form: $f_0=\tanh(\epsilon/2T)$.

 Canonical variables $\mu,\theta_p$ fulfill the Hamilton equations:
\begin{equation}\label{HamiltonEqs}
\dot{\theta_p}=\frac{\partial \varepsilon_n}{\partial \mu};\;\;\;
\dot{\mu}=-\frac{\partial \varepsilon_n}{\partial \theta_p}.
 \end{equation}
With the help of Hamilton equations
 the kinetic equation
(\ref{kinetic}) can be rewritten in the following
\begin{equation}\label{kinetic1}
 \frac{\partial f}{\partial t}+\{\varepsilon_n, f\}=St(f),
\end{equation}
where we use the Poisson bracket
 operator:
  $$
 \{\varepsilon_n,f\}=\frac{\partial f}{\partial \theta_p}\frac{\partial \varepsilon_n}{\partial
 \mu}-\frac{\partial f}{\partial \mu}\frac{\partial \varepsilon_n}{\partial
 \theta_p}.
 $$
 Using the Hamilton
 equations it is easy to see that $\{\varepsilon_n, f_0\}=0$ as well as
 $$ \left\{\varepsilon_n,\frac{d f_0}{d\varepsilon}\right\}=0,
 $$
 Thus if $f=f_0$ the only term that survives in kinetic equation is
  $$
 \frac{\partial f_0}{\partial t}=\frac{d f_0}{d \varepsilon}\frac{\partial \varepsilon_n}{\partial
 t}.
 $$
 Hence for the first order correction to the distribution
function $f=f_0+f_1$ we obtain the equation:
 \begin{equation}\label{Correction1}
 \frac{\partial f_1}{\partial t}+\{\varepsilon_n,f_1\}-\frac{f_1}{\tau}=
 -\frac{d f_0}{d \varepsilon}\frac{\partial \varepsilon_n}{\partial
 t}.
 \end{equation}
This kinetic equation together with the expression for the
dissipation losses (\ref{DissLossSuscept}) are the basic equations
which we will use to analyze the paramagnetic response of vortex
cores.


\section{Results} \label{Results}

\subsection{Quasiparticle spectrum.} \label{Results:spectra1}

 In zero magnetic field the BdG equations (\ref{BdG-U},\ref{BdG-V}) pertain the axial symmetry of wave functions which
 is described by the symmetry generator:
\begin{equation}\label{Qqp}
\hat Q_{qp}=\hat J_z-(M/2)I\tau_3,
\end{equation}
 where $\tau_3$ is a Nambu spin.
Then the quantum number which characterizes quasiparticle spectrum
and enters the Caroli - de Gennes Matricon expression is the
eigenvalue of the symmetry generator $\hat Q_{qp}\Psi=\mu \Psi$.

\begin{figure}[h!]
 \centerline{\includegraphics[width=1.0\linewidth]{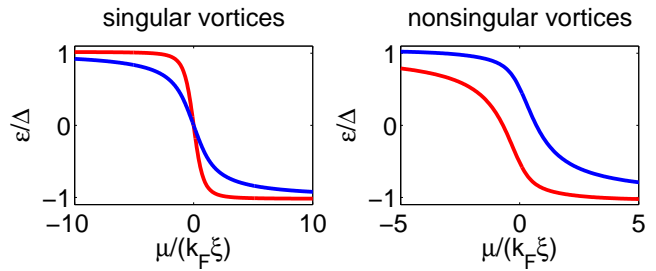}}
 \caption{\label{AnBr} Anomalous energy branches $\varepsilon=\varepsilon (\mu)$ for (a) singular
 $o$ and $u$ vortices; (b) nonsingular vortices. }
   \end{figure}

In general due to the removed spin degeneracy the spectra of all
singly quantized vortices in $^3$He B consist of two different
anomalous branches crossing the Fermi level\cite{SilaevHe3}. The
examples of anomalous branches for the singular and nonsingular
vortices obtained by numerical solution of BdG equations
(\ref{BdG-U},\ref{BdG-V}) are shown in Figs.(\ref{AnBr}a) and
(\ref{AnBr}b) correspondingly.

For small energies near the Fermi level an analytical treatment of
energy spectrum of Eqs.(\ref{BdG-U},\ref{BdG-V}) is
possible\cite{SilaevHe3}.  For singular $o$ and $u$ vortices the
anomalous branches are similar to the standard Caroli-de Gennes
-Matricon ones and intersect
 the Fermi level at zero angular momentum yet with
 different slopes corresponding to different spin states:
\begin{equation}\label{ou-vortex-Spectrum}
  \varepsilon(\mu,\chi)=-\hbar\omega_{\chi} \mu,
 \end{equation}
 $\hbar\omega_{\chi}\sim\Delta_0/(k_F\xi)$ and $\chi=\pm 1$
 corresponds to the different spin states. The difference in slopes of anomalous branches
   is determined by the asymmetry of amplitudes $C_{1,-1}, C_{-1,1}$ and $C_{0,0}$
   inside vortex core. Further we will assume that this asymmetry is small to neglect the
   spin degeneracy of anomalous branches for singular vortices and
   put $\omega_1=\omega_{-1}=\omega_v$.

 On the contrary the spectral branches of nonsingular vortices $v$, $w$ and $uvw$ intersect the Fermi
 level at finite angular momenta [see Fig.(\ref{AnBr})]:
\begin{equation}\label{vw-A-phase-Spectrum}
  \varepsilon(\mu,q_z, \chi)=-\hbar\omega_\chi \mu+\gamma_1  q_z+\chi \gamma_2,
 \end{equation}
 where $\gamma_{1,2}$ as well as $\omega_\chi$ are even functions of $q_z=k_z/k_F$ and $\mu$.
Such requirements provide spectrum symmetry
$\varepsilon(\mu,q_z,\chi)=-\varepsilon(-\mu,-q_z,-\chi)$
corresponding to the general invariance of BdG equations
(\ref{QuasiSymmetry}).
For the nonsingular vortices it is essential that the last term in
Eq.(\ref{vw-A-phase-Spectrum}) corresponding to the spin splitting
of energy branches can be rather large $\gamma_2\sim \Delta_0$.
Contrary to the case of singular vortices when calculating the
spectrum transformation due to the applied magnetic field we will
assume that these energy branches are weakly interacting, i.e. the
splitting is much larger than the Zeeman terms in Eqs.
(\ref{BdG-U},\ref{BdG-V}).

Now let us suppose that there is a magnetic field applied to the
system. In case when the magnetic field has a component ${\bf
H}_\perp$ which is perpendicular to the vortex axis ${\bf z}$ the
Zeeman term does not commute with the operator $\hat Q_{qp}$ given
by (\ref{Qqp}) and its eigenvalue $\mu$ is no more a good quantum
number. Therefore the quasiclassical spectrum of BdG equations
(\ref{BdG-U},\ref{BdG-V}) should depend not only on $\mu$ but also
on the conjugated angle variable which in our case coincides with
$\theta_p$ such that $[\hat Q_{qp}, \theta_p]=i$. Note that the
angular dependence of the spectrum is determined by the component
of the magnetic field ${\bf H}_\perp$ perpendicular to the vortex
axis since the $H_z$ component does not destroy the axial
symmetry.

 In general the BdG system (\ref{BdG-U},\ref{BdG-V}) can be
solved only numerically. However the general properties of
spectrum transformation in external magnetic field can be derived
from the symmetry properties of the different types of vortices.
We will consider the expansion of energy spectrum by powers of
magnetic field ${\bf H}$ assuming that the Zeeman terms are much
smaller than the spacing of quasiclassical levels determined by
the energy scale $\Delta$.

 The spectrum should obey the
symmetry relations (\ref{QuasiSymmetry},\ref{QuasiSymmetryH}) and
be invariant under the simultaneous rotation of coordinate axes
and magnetic field around the vortex axis ${\bf z}$. Then up to
the first order in $H_\perp$ we can write the spectrum perturbed
by the magnetic field:
 \begin{equation}\label{SpectrumGeneral}
 \varepsilon(\mu, \theta_p, \chi)=\varepsilon_0 + \alpha_1 ({\bf q}\cdot {\bf
 H}_\perp)+
  \alpha_2 ({\bf z}\cdot [{\bf
 q}\times {\bf H}_\perp]),
 \end{equation}
 where ${\bf q}={\bf k}/k_F$.
  The first term in this expression describes the axially symmetric part of the spectrum.
  Although it can depend on magnetic field for our further consideration this dependence
 will be of no importance. Thus we assume that the first term in
 Eq.(\ref{SpectrumGeneral}) corresponds to the spectrum without
 external magnetic field given by Eqs.(\ref{ou-vortex-Spectrum},\ref{vw-A-phase-Spectrum}).
  The remaining terms correspond to the spectrum
 perturbation due to the field introduction and the coefficients $\alpha_{1,2}$
 do not depend on the angle $\theta_p$. It is easy to check that this expression is
 invariant under the simultaneous rotation of coordinate axes and magnetic
 field around the vortex axis ${\bf z}$.
 The coefficients $\alpha_{1,2}$ should be expressed
 through the characteristics of the order parameter
 distribution inside vortex core as well as the characteristics of the unperturbed
 quasiparticle wave function.

We start with the spectrum of singular vortices which in zero
magnetic field is given by the Eq.(\ref{ou-vortex-Spectrum}) with
$\omega_1=\omega_{-1}=\omega_v$. We assume that this spectrum is
degenerate by spin, therefore the transformation should be linear
in Zeeman shift which lifts the spin degeneracy.
 The only possibility to fulfill
the requirements above is to put $\alpha_{i}=\chi \beta_{si}$,
where $\beta_{si}$ are scalar coefficients, which do not depend on
$\chi$, $\theta_p$, and are even in $\mu$ and $q_z$.

Now let us turn to the case of nonsingular vortices with lifted
spin degeneracy even in non-perturbed spectrum
(\ref{vw-A-phase-Spectrum}).
 It occurs that we can determine the general properties of the spectrum
 (\ref{SpectrumGeneral}) from the symmetry requirements.
  Indeed, the spectrum should be invariant under the time $T$ and
 spatial $P$ inversions and also under the rotation of the
 coordinate system $U_2$. Also we note that the energy perturbation should be quadratic
 in magnetic field due to the spin splitting of the spectral
 branches (\ref{vw-A-phase-Spectrum}) and the general spectrum
 symmetry (\ref{QuasiSymmetryH}) with respect to the total sign of
 magnetic field ${\bf H}$. Thus both the coefficients $\alpha_{1,2}$
 should be proportional to the projection of the
 magnetic field on the vortex axis $H_z=({\bf H}\cdot {\bf z})$.

 At first let us assume that the coefficients $\alpha_{1,2}$ do not depend on spin.
 Then the only possibility to satisfy the symmetry requirements
 above is to choose
 \begin{equation}\label{alpha1}
 \alpha_{1,2}\sim \alpha_p H_z
 \end{equation}
 where $\alpha_p$ is a pseudoscalar given by the
 Eq.(\ref{pseudoscalar}):
\begin{equation}\label{DSymmPseudo}
 P,T,U_2(\alpha_p)=-\alpha_p.
 \end{equation}

 If we neglect the asymmetry of B phase components within the cores
 of nonsingular vortices and put $C_{-1,1}=C_{1,-1}=C_{0,0}$ then
 from the definition of the pseudscalar (\ref{pseudoscalarUVW}) we
 immediately obtain that the spectrum of the $w$ and $uvw$ vortices then the
 symmetry allows that both the $\alpha_{1,2} \neq 0$.

 As for the $v$ vortex the pseudoscalar  defined by Eq. (\ref{pseudoscalarUVW})
 is obviously zero in this case. However $\alpha_{1,2}$ can be nonzero
 even for the $v$ vortex if we assume that they can depend of
 quasiparticle spin quantum number $\chi$. Indeed, the coefficient $\gamma_2$
 in the spectrum (\ref{vw-A-phase-Spectrum}) can be modified to
 include the magnetic field dependent terms as follows:
 $$
 \tilde{\gamma}_2=\gamma_2 \left(1+\beta_1 q_z H_z ({\bf q}\cdot {\bf H}_\perp)+
 \beta_2 q_zH_z ({\bf z}\cdot [{\bf H}_\perp\times {\bf q}])\right),
 $$
 where $\beta_{1,2}$ are ordinary scalars. It is easy to check
 that the spectrum (\ref{vw-A-phase-Spectrum}) with modified $\tilde{\gamma}_2$
 satisfies the same discrete symmetries as the non-perturbed
 spectrum. Thus the coefficients in expansion
 (\ref{SpectrumGeneral}) can be chosen as $\alpha_1=\chi\gamma_2\beta_1
 q_zH_z$ and $\alpha_2=\chi\gamma_2\beta_2 q_z$.

 Thus we can conclude that the quasiparticle spectrum
 of singular and nonsingular vortices has the same form of Eq. (\ref{SpectrumGeneral})
 with $\alpha_{1,2}\neq 0$ for all types of vortices.
 However, the magnitude of coefficients $\alpha_{1,2}\neq 0$ is
 different for singular and nonsingular vortices. Indeed, in the
 former case the initial spectrum is assumed spin degenerate, therefore
 energy perturbation is of the first order in magnetic field, thus
 $\alpha_{1,2}\sim \mu_B$. On the other hand the spin degeneracy of the spectrum
 of nonsingular vortices is lifted even at zero magnetic field.
 Therefore the energy perturbation is nonzero only in the second
 order of magnetic field and the coefficients in (\ref{SpectrumGeneral})
 are proportional to $\alpha_{1,2}\sim (\mu_B^2/\Delta_{s}) H_z$,
 where $\Delta_s$ denotes the energy of initial split of energy branches.
 In general, the splitting can be rather large $\Delta_s\sim
 \Delta_0$ (see Ref.\onlinecite{SilaevHe3}) therefore we obtain that for nonsingular vortices the
 energy perturbation is smaller by the factor $\mu_B H_z/\Delta_0$
 than for the singular vortices.

  In general the spectrum (\ref{SpectrumGeneral}) contains
 the terms which break angular symmetry and depend on the external magnetic field.
 Hence by changing the magnetic
 field ${\bf H}$ it is possible to excite the magnetic dipole
 transitions of quasiparticles between the neighboring Caroli - de Gennes Matricon
 levels, which should lead to the resonant energy absorption for a
 definite frequency of magnetic field oscillations.

  Finally in this section we should note that the possible rotation
 of spin quantization axes with respect to the orbital ones given
 by the Eq. (\ref{RotateOP}) can be easily taken into account in
 the above argument. The only thing we need is to transform the
 quasiparticle wave functions
 as follows:
 \begin{equation}\label{SpinRotate}
  (\tilde{U},\tilde{V})=\exp (i ({\bf \sigma}\cdot {\bf n})
 \varphi) (U,V),
 \end{equation}
 where ${\bf n}$ is a rotation axis and $\varphi$ is a rotation angle
 which parameterize the rotation matrix:
 $$
 (\hat R)_{ik}=\delta_{ik}+(n_in_k-\delta_{ik})
 (1-\cos\varphi)-e_{ikl} n_l \sin\varphi.
 $$

  It is straightforward to check
 that such transformation (\ref{SpinRotate}) if applied to Eqs.(\ref{BdG-U},\ref{BdG-V})
 makes the spin axes of the order parameter coincide with the
 orbital ones. But simultaneously it leads to the effective
 rotation of magnetic field
\begin{equation}\label{RotatedField}
\tilde{{\bf H}}=\hat R {\bf H}.
\end{equation}
  Note that the matrix
 $\hat R$ does not depend on the angle $\theta_p$. Therefore
  the only change that should be done in the above
 consideration to take into account the rotation of spin axes is
 to replace everywhere the magnetic field by the rotated one
 (\ref{RotatedField}).

 \subsection{Numerical solution of BdG equations.} \label{Results:spectra2}

  To confirm the general argument above we solve numerically the
 set of quasiclassical BdG equations (\ref{BdG-U},\ref{BdG-V}) to obtain the spectrum
 $\varepsilon=\varepsilon (\mu,\theta_p)$.

 We consider the model form of the vortex core
 such that the components corresponding to the $B$ phase are equal
 $C_{1,-1}=C_{-1,1}=C_{0,0}= C_B$ and only an additional $A$ phase component
 is present inside vortex core.
 Then the singular part of gap operator in $(s,\theta_p)$ representation is
 \begin{equation}\label{modelDeltaB0-st}
 \hat{\Delta}_B=C_{B}\frac{s-ib}{\sqrt{s^2+b^2}}
 \hat D,
 \end{equation}
 where
 $$
 \hat D=-q_z\hat\sigma_xe^{i\theta_p}+q_\perp\hat\sigma_ze^{i(1-\hat\sigma_z)\theta_p},
 $$
  ${\bf q}={\bf k_F}/k_F$ and $q_\perp=k_\perp/k_F$,
  $q_z=k_z/k_F$.
 The nonsingular part of gap function $\hat{\Delta}_{A}$ is is given by
 \begin{equation}\label{modelDeltaA0-st}
 \hat{\Delta}_{A}=-C_{A}q_\perp e^{i\theta_p}\hat\sigma_x.
 \end{equation}

\begin{figure}[h!]
 \centerline{\includegraphics[width=1.0\linewidth]{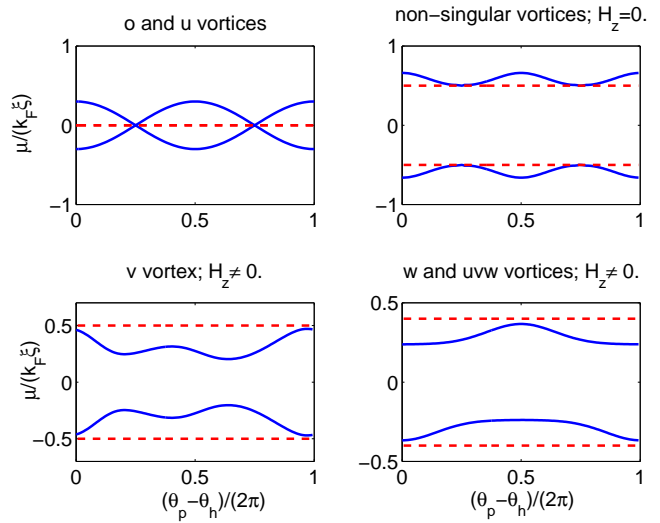}}
 \caption{\label{Q-teta} Isoenergetic lines $\mu(\theta_p)$
  corresponding to the  different types of vortices in He3-B. Dash
  lines show $\mu(\theta_p)$ in zero magnetic field and solid
  lines correspond to the modification of  $\mu(\theta_p)$ by the
  applied field.}
   \end{figure}

 In Fig.(\ref{Q-teta}) we show the
 isoenergetic lines on the plane $\mu,\theta_p$ corresponding to the
 zero energy $\varepsilon=0$ for several generic cases: (i) singular $o$ and $u$ vortices;
 (ii) nonsingular vortices for $H_z=0$; (iii) nonsingular $v$ vortex
 for $H_z\neq 0$; (iv) $w$ and $uvw$ vortex for
 $H_z\neq 0$.

 To understand the numerical results shown in Fig.(\ref{Q-teta})
 let us consider the expressions for isoenergetic lines $\mu=\mu(\theta_p,\varepsilon)$
 which can be derived from the general expression for quasiparticle spectrum
  (\ref{SpectrumGeneral}). For singular
  vortices we obtain
 \begin{equation}\label{SingIsoEn}
 \mu (\theta_p)=\chi \tilde{\beta}_1 \cos(\theta_p-\theta_h)+\chi \tilde{\beta}_2
 \sin(\theta_p-\theta_h),\end{equation}
   for nonsingular $v$ vortices
\begin{equation}\label{VIsoEn}
\mu (\theta_p)=\mu_0+
\chi\tilde{\alpha}_{1v}\cos(\theta_p-\theta_h) +
\chi\tilde{\alpha}_{2v} \sin(\theta_p-\theta_h),
\end{equation}
and for for nonsingular $w$ and $uvw$ vortices
\begin{equation}\label{WIsoEn}
\mu (\theta_p)=\mu_0+ \tilde{\alpha}_{1w}\cos(\theta_p-\theta_h) +
\tilde{\alpha}_{2w} \sin(\theta_p-\theta_h),
\end{equation}
where we introduce the angle $\theta_h$ characterizing the
direction of perpendicular component of magnetic field ${\bf
H}_\perp= H_\perp (\cos\theta_h, \sin\theta_h)$.

 Then from the Fig.(\ref{Q-teta}) we can see that for the model
 vortex core (\ref{modelDeltaB0-st},\ref{modelDeltaA0-st}) the
 isoenergetic lines are indeed given by Eqs.(\ref{SingIsoEn},\ref{VIsoEn},\ref{WIsoEn})
 with $\tilde{\beta}_1\neq 0$ and $\tilde{\beta}_2=0$ for singular
 vortices; $\tilde{\alpha}_{1v},\tilde{\alpha}_{2v}\neq 0$ for nonsingular $v$ vortex;
 $\tilde{\alpha}_{2w}\neq 0$ and $\tilde{\alpha}_{1w}=0$ for nonsingular
 $w$ and $uvw$ vortices. In case if $H_z=0$ we have that all the
 coefficients are zero
 $\tilde{\alpha}_{1v},\tilde{\alpha}_{2v},\tilde{\alpha}_{1w},\tilde{\alpha}_{2w} = 0$
 and the spectrum transformation should be of the next order in ${\bf H}_\perp$,
 i.e. proportional to $({\bf q}\cdot {\bf H_\perp})^2\sim
 H_\perp^2\cos(2(\theta_p-\theta_h))$, which is also demonstrated in
 Fig.\ref{Q-teta} (see the plot for non-singular vortices; $H_z=0$).

 In principle the numerical calculations described above allow to
 consider the spectrum transformation for the arbitrary values of
 Zeeman shift. However, the qualitative expression
  (\ref{SpectrumGeneral})
 have an advantage that they also take into account the
 self-consistent transformation of gap function
 in magnetic field $\Delta=\Delta({\bf H})$. If this transformation is small it does not change the vortex symmetry
 therefore the general form of the energy spectrum conserves yet
 with modified coefficients  $\alpha_{1,2}$ in Eq.(\ref{SpectrumGeneral}).
 In our further considerations we will take these coefficients
 as phenomenological constants since their particular values does
 not affect the results qualitatively.

\subsection{Magnetic susceptibility and energy dissipation.}
\label{Results:magn}

Now having in hand the expression for quasiparticle energy
(\ref{SpectrumGeneral})
 it is then straightforward to calculate the non-stationary magnetization of vortex cores with
 the help if kinetic equation (\ref{kinetic}) and the general expression for
 the magnetization (\ref{MagnQP}).

 We will calculate
 the response magnetization of a single vortex using a unified
 expression for the spectrum (\ref{SpectrumGeneral}) assuming
 that only the component of magnetic field perpendicular to the
 vortex line varies in time. Also without loss of generality we
 put $\alpha_1\neq 0$ and $\alpha_2=0$, which can always be done
 by rotating the coordinate frame around the vortex axis by an
 appropriate angle.

  At first we will assume that
   an ac component of magnetic field is the one perpendicular
  to the vortex line.
  The component $H_z$ is assumed in general non-zero time independent.
  Also for the beginning the perpendicular component is taken polarized along the ${\bf x}$ axis
  ${\bf H}_\perp =H_x e^{i\omega_{rf} t} {\bf x}$. Then solving the
  kinetic equation (see Appendix B) and calculating the magnetization with the help
  of Eq.(\ref{MagnQPst}) we obtain the following form of the
  magnetic susceptibility tensor $\chi_{xx}=\chi_{yy}= \chi_\parallel$
  and $\chi_{xy}=-\chi_{yx}=\chi_\perp$, where
 \begin{equation}\label{SuscPar}
  \chi_\parallel=  \int_{-k_F}^{k_F} \frac{dk_z}{\omega_v}
  \frac{\alpha(\omega_{rf}+i/\tau)\omega_{rf}}{\omega_v^2-(\omega_{rf}+i/\tau)^2}
  \end{equation}
 \begin{equation}\label{SuscPerp}
 \chi_\perp= i\int_{-k_F}^{k_F} \frac{dk_z}{\omega_v}
  \frac{\alpha\omega_v\omega_{rf}}{\omega_v^2-(\omega_{rf}+i/\tau)^2}.
  \end{equation}
  Here we have denoted
  $\omega_v=-\hbar^{-1}\partial\varepsilon_n/\partial \mu$
  which is the frequency corresponding to the
 interlevel energy spacing. Further we will assume that this
 spacing is the same for all anomalous branches, i.e. $\omega_v$
 does not depend on the index $n$.
  Here we have used the relation $(\hbar\omega_v)\partial f_0/\partial \varepsilon=- \partial f_0/\partial \mu$
   and integrated over $\mu$ assuming for simplicity that  $\omega_v$ and $\alpha$
    do not depend on $\mu$ in order to perform the integration over
  $\mu$. We have denoted  $\alpha=N_v\alpha^2_1/(8\pi^2)$, where
 the overall factor $N_v$ is a vortex density.

The magnetic susceptibility defined by the
Eqs.(\ref{SuscPar},\ref{SuscPerp}) has resonances at the
frequencies $\omega_{rf}=\pm \omega_v$. If the interlevel spacing
$\omega_v$ depends on $k_z$ [such as in Eq.(\ref{omega(kz)})] the
resonance peak transforms into the band with the absorption edge
at $\omega_{rf}=\omega_0={\rm min} |\omega_v (k_z)|$. In this case
for the large enough relaxation times $\tau\gg \omega_0$ the
susceptibility is also peaked at $\omega_{rf}=\pm \omega_0$
behaving as
  $$
  \chi_\perp,\chi_\parallel\sim
  \frac{1}{\tau\sqrt{\omega_{rf}^2-\omega_0^2}}.
  $$

 The
frequency independent magnitude of susceptibility can be estimated
as
 $$
 \chi_\perp,\chi_\parallel\sim \chi_n \xi^2 N_v
 $$
 for singular vortices and
 $$
  \chi_\perp,\chi_\parallel\sim \chi_n \left(\frac{\mu_B
  H_z}{\Delta_s}\right)^2\xi^2 N_v
  $$
for nonsingular vortices, where $\chi_n\sim \nu_0 \mu_B^2$ is a
susceptibility of normal phase and $\Delta_s$ is the energy
splitting of anomalous branches which can be taken of the order of
bulk value of energy gap $\Delta$.

 The resonant frequency of
paramagnetic response where the absorption maximum takes place is
selective to the polarization of external magnetic field. To
demonstrate this let us consider two limiting cases: the linear
polarization ${\bf H}=H_x e^{i\omega_{rf} t} {\bf x}$ and the
circular one ${\bf H}=H_\perp e^{i\omega_{rf} t} ({\bf x}+iP{\bf
y})$, where $P=\pm 1$ determines the direction of field rotation.
In former case the dissipation determined by the
Eq.(\ref{DissLossSuscept}) has the form
 $$
 Q_\varepsilon=\frac{\omega_{rf} H_x^2}{2} Im \chi_\parallel,
 $$
 where
 $$
 Im \chi_\parallel =
$$ $$
 \int_{-k_F}^{k_F} dk_z
 \left[\frac{\alpha\omega_v\omega_{rf} \tau}{(\omega_v-\omega_{rf})^2\tau^2+1}-
 \frac{\alpha\omega_v\omega_{rf} \tau}{(\omega_v+\omega_{rf})^2\tau^2+1}
 \right].
 $$
 So the resonant energy absorption takes place both at $\omega_{rf}=\omega_v$
 and $\omega_{rf}=-\omega_v$.

 In case of the circular polarization the dissipation rate is
 given by
 $$
 Q_\varepsilon=\frac{\omega_{rf} H_\perp^2}{2} Im (\chi_\parallel+i\chi_\perp),
 $$
 where
  $$
  Im (\chi_\parallel+i\chi_\perp) = \int_{-k_F}^{k_F} dk_z
  \frac{\alpha\omega_v\omega_{rf} \tau}{(\omega_v+P\omega_{rf})^2\tau^2+1}.
  $$
  For the circular polarization the resonant frequency in the expression above depends on
  the direction of magnetic field rotation $\omega_{rf}=P\omega_v$.

 Thus we see that the paramagnetic susceptibility
 as well as the dissipation losses in the system have the
 resonances defined in general by the interlevel spacing in the
 spectrum of vortex core fermion states
 (\ref{ou-vortex-Spectrum},\ref{vw-A-phase-Spectrum}).

 Interestingly, if we assume that the projection of magnetic field
 onto the vortex line is  also time-dependent $H_z = H^0_z+ \tilde{H}_z\cos(\omega_{rf}
 t+\delta)$ then besides the described above resonance at the main frequency $\omega_{rf}=\pm\hbar^{-1}\partial\varepsilon/\partial \mu$
 there can also appear resonances at higher frequencies $\omega_{rf}= n
 \omega_v$ and at fractional frequencies $\omega_{rf}=
 \omega_v/n$, where $n$ is an integer number. This situation can naturally be realized
 in the experiment since it is not the real magnetic field which
 determines magnetic response of vortices but a rotated one according to the Eq.(\ref{RotatedField}).
 In general the rotation matrix $\hat R$ is spatially dependent
 hence the effective field $\hat R {\bf H}$ has different directions at different points of the superfluid.
 To show the possibility of the additional resonances let us
 note that the expression for spectrum (\ref{SpectrumGeneral})
 is in general nonlinear in ${\bf H}$. In particular, for the
 nonsingular vortices the coefficients in (\ref{SpectrumGeneral}) are proportional to $H_z$ and therefore depend on time as
 $\alpha_{1,2}=\alpha_{1,2}^0+  \tilde{\alpha}_{1,2}\cos(\omega_{rf}
 t+\delta)$. This additional modulation of the coefficients
 leads to the appearance of $2\omega_{rf}$ frequency terms in the energy spectrum
 (\ref{SpectrumGeneral}). Then solving the kinetic equation in
 standard way described in Appendix (\ref{App2}) yields the
 resonances at $\omega_{rf}=\pm \omega_v/2$.
  In order to obtain the
 resonances at other frequencies it is necessary to consider the
 higher terms in spectrum expansion by the powers of the magnetic
 field ${\bf H}$.


\section{Summary} \label{sec:summary}

  To sum up, we have investigated the spectrum of bound fermion states on vortices in $^3$He B
 modified by an external magnetic field. We have developed a general approach to study the spectrum perturbation
 based on symmetry grounds. It allowed us to determine  qualitatively up to the constants of the order unity
 the form of bound states spectrum for different types of vortices in $^3$He
 B. An important advantage of this phenomenological approach to the
 spectral problem is a possibility to take into account the
 modification of the order parameter by the external field without
  extensive numerical calculations.

 We also consider the paramagnetic susceptibility of fermionic ensemble bound within
   vortex cores. We have shown quite generally that it is the fermionic magnetization which
  determines the energy losses in ac external magnetic field driving the system out of
  equilibrium. It occurs that due to the coupling of orbital and
  spin quasiparticle degrees of freedom the ac magnetic field
  induces transitions of bound fermions between different energy
  levels in a ladder of Caroli - de Gennes Matricon spectrum
  (\ref{CdGMspectrum}). Consequently the paramagnetic susceptibility and
  energy absorption have resonance which occurs when the frequency
 of an ac external magnetic field equals the interlevel spacing.
 Due to the broken time inversion symmetry of vortex state the
 resonant behaviour of energy dissipation depends on the
 polarization of ac magnetic field. In particular for a circularly
 polarized magnetic field rotating over the vortex axis the presence of resonance
 depends on the relation between vortex winding direction and the direction
 of field rotation.

  Although the resonant absorption occurs at the same frequency for
 singular and nonsingular vortices the dissipation rate should be quite different in this two
 cases. Being proportional to $Q_\varepsilon\sim \omega_{rf}\mu_B H_\perp^2$ for the $o$ and $u$ vortices it
 is much less for the nonsingular vortices when $Q_\varepsilon\sim\omega_{rf} \mu_B H_\perp^2
 (\mu_B H/\Delta_0)^2$ since the magnetic field is assumed to
 much smaller than the spin depairing one so that $\mu_B H\ll \Delta_0$.

 Due to the resonant behaviour of paramagnetic susceptibility at the frequency
 of interlevel transitions within the Caroli - de Gennes Matricon spectrum we
  can conclude that measuring of a resonant magnetic
 susceptibility of vortex cores in $^3$He B can provide a
 tool to study the discrete nature of bound fermions in
 vortex core. Also the difference in energy absorption rates for singular and non-singular vortices
 can provide an evidence for the particular type of vortices realizing in $^3$He under different
 experimental conditions.




 \section{Acknowledgements}
It is our pleasure to thank G.E. Volovik,  A.S.
  Mel'nikov and N.B. Kopnin for numerous stimulating discussions.
 This work was supported, in part, by "Dynasty" Foundation, Russian Foundation for Basic Research,
 by Programs of RAS "Quantum Physics of Condensed Matter" and "Strongly
  correlated electrons in semiconductors, metals, superconductors and magnetic
  materials".


\appendix

\label{App1}
\section{Symmetries of quasiparticle spectrum.}

We are going to prove the general symmetries
(\ref{QuasiSymmetry},\ref{QuasiSymmetryH}) of the spectrum of BdG
system (\ref{BdG-U},\ref{BdG-V}). At first let us consider the
symmetry (\ref{QuasiSymmetry}).
 Note that
the coordinates in real space are related to the coordinates $s,b$
as follows:
 $$
 x=s\cos\theta_p+b\sin\theta_p
 $$
 $$
 y=s\sin\theta_p-b\cos\theta_p.
 $$
 Thus the transformation of $s,b,\theta_p$ to $-s,-b,\theta_p+\pi$
 does not change the coordinates $x,y$. It means that the coordinate part
 of function $\hat\Delta_{\bf k}$ remains intact. At the same time changing
 $k_z, \theta_p$ by $-k_z, \theta_p+\pi$ leads to the reverse of
 momentum direction. Then the total gap function
 changes its sign under the transformation
 $\hat\Delta_{\bf k} (s,\theta_p,k_z,b)=-\hat\Delta_{\bf k}
 (-s,\theta_p+\pi,-k_z,-b)$.  Note that the matrix $\hat \Delta_k$ does not contain the
Pauli matrix $\hat\sigma_y$ therefore $\hat \Delta_k^+=\hat
\Delta_k^*$. Then the complex conjugation of transformed BdG
equations yields
 \begin{equation}\label{BdG-V1}
 -i\frac{\hbar k_\perp}{m} \frac{\partial}{\partial s}
 V^*+\hat{\Delta}_{\bf k}U^*=\left(-\varepsilon-\hat P\right)V^*,
 \end{equation}
\begin{equation}\label{BdG-U1}
 i\frac{\hbar k_\perp}{m} \frac{\partial}{\partial s}
 U^*+\hat\Delta^+_{\bf k} V^*=\left(-\varepsilon+\hat
 P^*\right)U^*.
 \end{equation}
Changing $\varepsilon$ by $-\varepsilon$ we obtain the system
coinciding with the initial set of Eqs.(\ref{BdG-U},\ref{BdG-V})
which proves the relation (\ref{QuasiSymmetry}).

Now let us consider the symmetry (\ref{QuasiSymmetryH}). We
suppose here that $\hat \Delta_k({\bf H}) =\hat \Delta_k (-{\bf
H})$.  Note that the matrix $\hat \Delta_k$ does not contain the
Pauli matrix $\hat\sigma_y$ therefore $\hat \Delta_k^+=\hat
\Delta_k^*$. Then the complex conjugated BdG equations has the
form
\begin{equation}\label{BdG-Vcc}
 -i\frac{\hbar k_\perp}{m} \frac{\partial}{\partial s}
 V^*+\hat{\Delta}_{\bf k}U^*=\left(\varepsilon+\hat P\right)V^*,
 \end{equation}
\begin{equation}\label{BdG-Ucc}
 i\frac{\hbar k_\perp}{m} \frac{\partial}{\partial s}
 U^*+\hat\Delta^+_{\bf k} V^*=\left(\varepsilon-\hat
 P^*\right)U^*.
 \end{equation}
 Then changing ${\bf H}$ by $-{\bf H}$ we obtain initial set of
 Eqs.(\ref{BdG-U},\ref{BdG-V}) which proves
 (\ref{QuasiSymmetryH}).

\section{Kinetic equation.}

\label{App2}

 In general the solution of kinetic equation (\ref{Correction1})
 can be found in the following form
\begin{equation}\label{1}
 f_1= G(\theta_p,t)\frac{d f_0}{d \varepsilon}.
\end{equation}
 Then
 $$
 \frac{\partial f_1}{\partial t}= \frac{\partial G}{\partial t}\frac{d f_0}{d
 \varepsilon}+G\frac{d^{2} f_0}{d
 \varepsilon^{2}}\dot{\varepsilon},
 $$
 where the second term can be neglected.
 Also,  $\{H,G\}=-\omega_v (\partial G/\partial\theta_p)$.
Therefore for the function $G(\theta_p,t)$ we obtain the following
equation:
 \begin{equation}\label{CorrectionG}
 \frac{\partial G}{\partial t}-\omega_v \frac{\partial G}{\partial\theta_p}-\frac{G}{\tau}
 = - \frac{\partial \varepsilon}{\partial t},
 \end{equation}
 where we denote $\hbar\omega_v=-(\partial \varepsilon/\partial \mu)$.

 Further we will consider the solution of kinetic equation (\ref{CorrectionG})
 when the quasiparticle spectrum is given by (\ref{SpectrumGeneral})
 with $\alpha_1\neq 0$, $\alpha_2= 0$  and ${\bf H}_\perp=H_x e^{i\omega_{rf} t} {\bf x}$.
 Then we have
 $$
 \frac{\partial \varepsilon}{\partial t}=
 i\omega_{rf} \alpha_1 q_x H_x e^{i\omega_{rf} t}.
 $$

 Let us search the solution of Eq.(\ref{CorrectionG}) in the
 following form:
\begin{equation}\label{AppGsing}
 G= (A \cos \theta_p + B  \sin
 \theta_p)e^{i\omega_{rf} t}.
 \end{equation}
 Substituting this form to the Eq. (\ref{CorrectionG}) for the
 coefficients $A$ and $B$ we obtain:
\begin{equation}\label{AppSingA}
 A=\omega_{rf} \alpha_1 \frac{(\omega_{rf}+i/\tau)H_x}{\omega_v^2-(\omega_{rf}+i/\tau)^2}
 \end{equation}

\begin{equation}\label{AppSingB}
 B=i\omega_{rf} \alpha_1 \frac{\omega_v H_x}{\omega_v^2-(\omega_{rf}+i/\tau)^2}.
 \end{equation}

\end{document}